\documentclass[journal]{IEEEtran}
\usepackage{amsmath}
\usepackage[utf8]{inputenc} 
\usepackage[T1]{fontenc}    
\usepackage{hyperref}       
\usepackage{url}            
\usepackage{booktabs}       
\usepackage{amsfonts}       
\usepackage{nicefrac}       
\usepackage{microtype}      
\usepackage{xcolor}         
\usepackage[pdftex]{graphicx}       
\usepackage{threeparttable}
\usepackage{amssymb}
\usepackage{amsmath}
\usepackage{tabularx}
\usepackage{latexsym}
\usepackage{makecell}
\usepackage{cite}
\usepackage{bbding}

\begin{document}
%
\title{HDR Video Reconstruction with a Large \\
	Dynamic Dataset in Raw and sRGB Domains}
%
%
%

\author{Huanjing Yue,~\IEEEmembership{Member,~IEEE}, Yubo Peng, Biting Yu, Xuanwu Yin, \\Zhenyu Zhou, 
Jingyu Yang,~\IEEEmembership{Senior~Member,~IEEE}
\thanks{This work was supported in part by the National Natural Science Foundation of China under Grant 62072331 and Grant  62231018.}
\thanks{H. Yue, Y. Peng, and J. Yang (corresponding author) are with the School of Electrical and Information Engineering, Tianjin University, China. }
\thanks{B. Yu, X. Yin, and Z. Zhou are individuals.}
}
\maketitle   

%


\begin{abstract}
High dynamic range (HDR) video reconstruction is attracting more and more attention due to the superior visual quality compared with those of low dynamic range (LDR) videos. The availability of LDR-HDR training pairs is essential for the HDR reconstruction quality. However, there are still no real LDR-HDR pairs for dynamic scenes due to the difficulty in capturing LDR-HDR frames simultaneously. In this work, we propose to utilize a staggered sensor to capture two alternate exposure images simultaneously, which are then fused into an HDR frame in both raw and sRGB domains.
In this way, we build a large scale LDR-HDR video dataset with  85 scenes and each scene contains 60 frames. Based on this dataset, we further propose a Raw-HDRNet, which utilizes the raw LDR frames as inputs. We propose a pyramid flow-guided deformation convolution to align neighboring frames. Experimental results demonstrate that 1) the proposed dataset can improve the HDR reconstruction performance on real scenes for three benchmark networks; 2) Compared with sRGB inputs, utilizing raw inputs can further improve the reconstruction quality and our proposed Raw-HDRNet is a strong baseline for raw HDR reconstruction. \textit{Our dataset and code will be released after the acceptance of this paper.}
\end{abstract}



\section{Introduction}

\IEEEPARstart{H}{igh} dynamic range (HDR) images (videos) are more attractive compared with the low dynamic range (LDR) ones since they can reveal the missing contents in the over-exposed and under-exposed regions. However, it is expensive or complicated to capture the HDR image or video directly~\cite{JanFroehlich2014CreatingCW,kronander2014unified,MichaelDTocci2011AVH}. The most popular method for HDR image reconstruction is merging multiple temporally varying exposure images (also known as bracketed exposure images) ~\cite{PaulDebevec1997RecoveringHD,NimaKhademiKalantari2017DeepHD,PradeepSen2012RobustPH,QingsenYan2019AttentionGuidedNF,QingsenYan2020DeepHI} by aligning neighboring images to the reference image and fusing the aligned results together.

\begin{figure}
	\centering
	\includegraphics[width=1\linewidth]{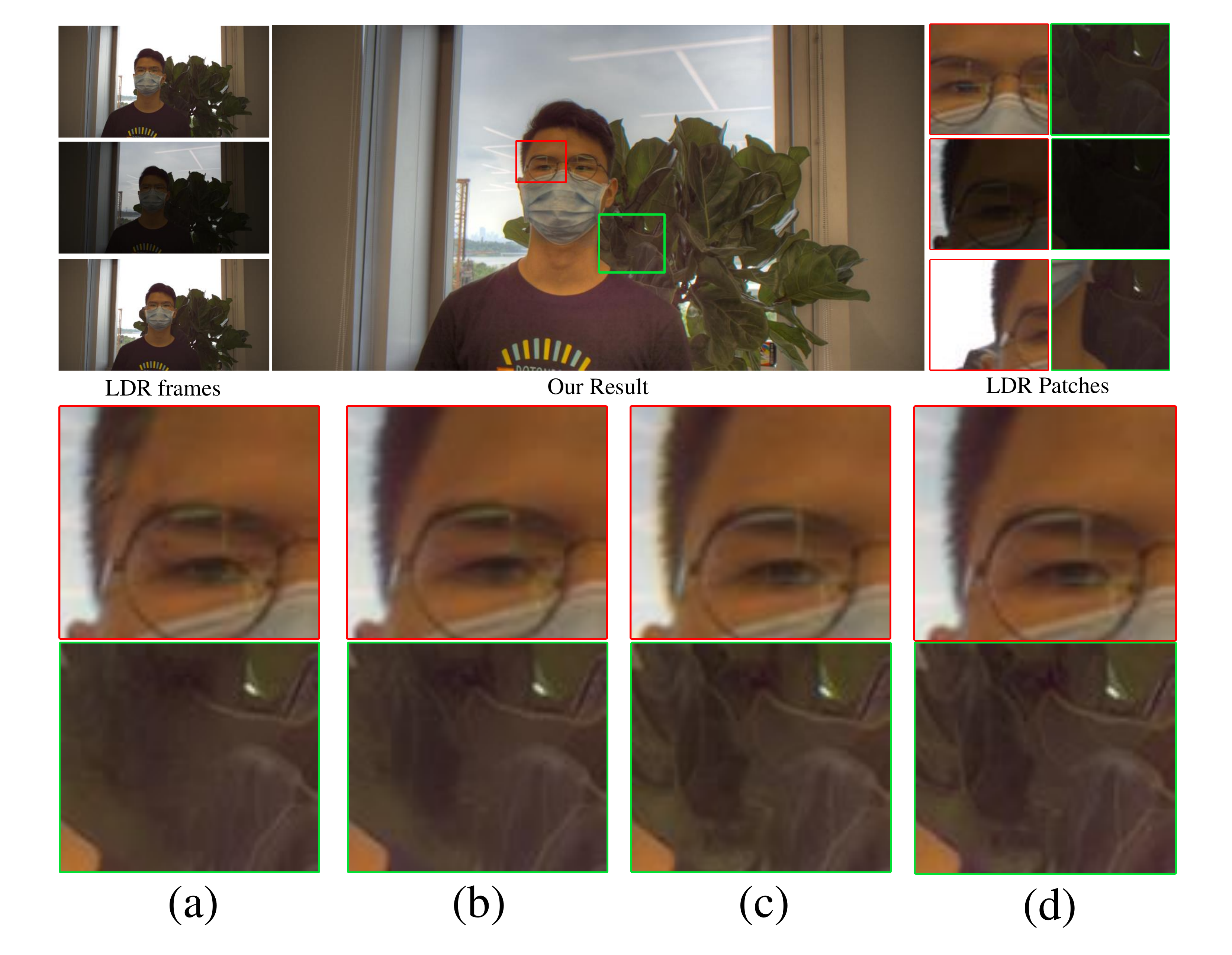}
	\caption{Comparison of the HDR reconstruction results on the real test set \cite{chen2021hdr} with different training sets.
		From left to right, the results are generated by \cite{chen2021hdr} trained with synthetic sRGB dataset (a),  \cite{chen2021hdr} trained with synthetic and our sRGB datasets (b), our Raw-HDRNet trained with  synthetic and our raw datasets (c) and the ground truth (d).}
	\label{fig:artifacts}
\end{figure}

Compared with HDR imaging, HDR videos are more attracting. Currently, there are mainly two methods to acquire HDR videos. One is hardware based, e.g., using the specialized hardware (\emph{e.g.}, beam splitter~\cite{JanFroehlich2014CreatingCW,kronander2014unified} or sensors with spatially varying exposures~\cite{VienGiaAn2017SingleshotHD,YitongJiang2021HDRVR}) to capture different exposure information, but this is not feasible for most cameras.
Another is software based, \textit{i.e.,} capturing video sequences with alternating exposures, which is feasible for most cameras, and then fuse them together~\cite{NimaKhademiKalantari2019DeepHV,SingBingKang2003HighDR}.
However, different from image HDR, which aligns the short and long exposures to the middle exposure image (i.e., reference image)~\cite{NimaKhademiKalantari2017DeepHD,ZhenLiu2021ADNetAD,QingsenYan2019AttentionGuidedNF}, the reference frame alternates between long-exposure and short-exposure frames in video HDR, which brings challenges to both alignment and fusion.
The reference frame needs complementary details from the neighbouring frames in the wrong exposed regions. However, the missed details prevent accurate alignment, which will lead to ghosting artifacts, as shown in Fig.~\ref{fig:artifacts} \footnote{All the HDR results in this paper are visualized after tone mapping with photomatix: https://fixthephoto.com/photomatix-free.html.}. Although many methods have been proposed to deal with the alignment problem, such as the patch-based~\cite{NimaKhademiKalantari2013PatchbasedHD} motion estimation,  optical flow  and  deformable convolution based alignment~\cite{chen2021hdr,NimaKhademiKalantari2019DeepHV}, they still cannot solve the problem well due to the missed details.

On the other hand, all of the learning based video HDR methods are trained on synthetic datasets due to the lacking of publicly available real-world LDR-HDR video dataset. The synthesized LDR sequences are usually obtained via image linearization and brightness scaling~\cite{chen2021hdr,NimaKhademiKalantari2019DeepHV}, which may suffer from two problems.
First, the gamma function used for inverse linearization is inconsistent with the real nonlinear camera response function, which cannot recover the real linearized image, such as the raw data.
Second, to simulate noise in the LDR image, they directly add Gaussian noise in the linearized LDR image, which is different from the complex noise distribution in real LDR frames. Due to these problems, the model trained by synthesized LDRs cannot work well in processing real LDR videos, as shown in Fig.~\ref{fig:artifacts}. Recently, Chen~\textit{et al.} proposed a real-world dataset for dynamic scenes~\cite{chen2021hdr}. However, there is no per-frame HDR ground truth for each video and only 76 LDR-HDR frame pairs are provided, which is too small for supervised training.

Besides, we observe that the real linearized frames, \textit{i.e.,} raw frames, have wider bit width, making the lost details in sRGB domain be discriminative again. Therefore, with raw frames as inputs is beneficial for alignment in under-exposed or over-exposed regions compared with the  pseudo linearized sRGB frames.
Directly processing image or video in the raw domain shows extraordinary advantages in denoising and super-resolution ~\cite{XiaohongLiu2021ExploitCR,XiangyuXu2019TowardsRS,HuanjingYue2020SupervisedRV}. However, HDR image (video) reconstruction in raw domain were rarely studied~\cite{PrashantChaudhari2019MergingISPMH,TianhongDai2021WaveletBasedNF,YanLiu2020AntiShakeHI} due to the lack of corresponding datasets.

Based on the above observations, we propose to construct a real-world LDR-HDR video dataset in both raw and sRGB domains. In order to generate frame-wise HDR ground truth, we propose to utilize the staggered readout mode in Xiaomi 10 Ultra phone. The staggered mode enables us to simultaneously read out long and short exposure images, which can be fused together to serve as the HDR ground truth frame, as shown in~ Fig. \ref{fig:dataset construction}. Then, we choose only one exposure image for each frame, and the switched long and short exposure frames build the input LDR sequence. Based on this dataset, we further propose an HDR video reconstruction network for raw inputs, termed as Raw-HDRNet.
In summary, the main contributions of this paper are as follows:

\begin{itemize}
	\item[$\bullet$] By taking advantage of the staggered readout mode, we construct a large-scale real-world LDR-HDR video dataset with per-frame ground truth. There are totally 85 scenes, each scene has 60 LDR-HDR frame pairs in both raw and sRGB formats. By retraining benchmark image and video HDR reconstruction networks on this dataset, the HDR reconstruction results are improved compared with the models trained with synthesized dataset.
\end{itemize}
\begin{itemize}
	\item[$\bullet$]
	We further propose a raw HDR reconstruction network (Raw-HDRNet) with alignment and fusion modules. For alignment, we propose a
	pyramid flow guided deformable (PFD) convlution, which can effectively dealing with large motions. Experimental results verify that 1) the raw LDR inputs are superior than the sRGB LDR inputs for HDR reconstruction, and 2) our proposed Raw-HDRNet is a strong baseline for raw HDR reconstruction. 
\end{itemize}


\section{Realted works}
\subsection{HDR Image Reconstruction}
Traditional single image HDR reconstruction expand the dynamic range of the LDR image by applying special mapping operators~\cite{FrancescoBanterle2006InverseTM,RafaelPachecoKovaleski2009HighqualityBE,BelenMasia2009EvaluationOR,ErikReinhard2002PhotographicTR,AllanGRempel2007Ldr2HdrOR}. In recent years, CNNs are used to inverse tone mapping and hallucinate missing details in the over-exposed regions~\cite{GabrielEilertsen2017HDRIR,YukiEndo2017DeepRT,SiyeongLee2018DeepRH,YuLunLiu2020SingleImageHR,MarcelSantanaSantos2020SingleIH}. Due to limited information in single image input, many methods proposed to
reconstruct the HDR image by merging bracketed LDR exposure sequence~\cite{KedeMa2017RobustMI,li2018densefuse,HuiLi2020FastMS}. However, directly merging multi-exposure images only works well in the static scenes but introducing ghosting artifacts in the moving regions~\cite{PaulDebevec1997RecoveringHD,mannbeing}. Traditional methods remove the wrong pixels in the motion region~\cite{ChulLee2014GhostFreeHD,TaeHyunOh2015RobustHD} or align the images before merging~\cite{OrazioGallo2015LocallyNR,JunHu2012ExposureSO,TomMertens2009ExposureFA,PradeepSen2012RobustPH}. Nowadays, CNNs are widely used in the HDR image reconstruction~\cite{ZhenLiu2021ADNetAD,ma2019deep,KRamPrabhakar2020TowardsPA,ShangzheWu2018DeepHD,YuzhenNiu2021HDRGANHI,QingsenYan2019AttentionGuidedNF,xu2020mef,deng2021deep}, where the spatial attention module and deformable convolution based alignment have greatly reduced the ghosting artifacts~\cite{ZhenLiu2021ADNetAD,QingsenYan2019AttentionGuidedNF}.
To enable the supervised training, the LDR-HDR image dataset, where the LDR images are real captured and the HDR image is obtained by fusing stable LDR sequence, is developed~\cite{NimaKhademiKalantari2017DeepHD}.

Besides these, there are also some methods focusing on raw image HDR since performing HDR fusion before the ISP process can reduce computing complexity. Some methods are based on specialized hardware, such as the spatial varying exposure~\cite{VienGiaAn2017SingleshotHD,ShreeKNayar2000HighDR} and beam splitter~\cite{JoelKronander2013UnifiedHR}. The other kind is aligning and fusing raw images in the raw domain~\cite{PrashantChaudhari2019MergingISPMH,TianhongDai2021WaveletBasedNF,YanLiu2020AntiShakeHI}. The work in \cite{PrashantChaudhari2019MergingISPMH} proposed to merging the HDR fusion and ISP process in a single network. However, its raw inputs are synthesized from the sRGB dataset \cite{NimaKhademiKalantari2017DeepHD}, which actually lacks wider bit depth. The work in~\cite{TianhongDai2021WaveletBasedNF} collected a raw image HDR dataset and proposed a network to construct HDR in the wavelet domain. However, it did not investigate special processing for raw inputs and powerful alignment module to remove ghosting artifacts.

\subsection{HDR Video Reconstruction}
HDR video reconstruction also includes two categories, the first is based on the specialized sensors or complicated devices to capture different exposures at the same time~\cite{InchangChoi2017ReconstructingIH,UgurCogalan2020DeepJD,JanFroehlich2014CreatingCW,YitongJiang2021HDRVR} and then fuse them together.
The other kind acquires alternative exposure sequences controlled by software and merges neighbouring frames to the reference frame. We would like to point out that this problem is more challenging than image HDR since the reference frame alternates between short and long exposure frames.
Many methods are proposed to solve the alignment problem, such as the optical flow based \cite{NimaKhademiKalantari2017DeepHD,SingBingKang2003HighDR}, block-based motion estimation \cite{StephenMangiat2010HighDR}, combination of patch-based synthesis and optical flow \cite{NimaKhademiKalantari2013PatchbasedHD}, and maximum posterior estimation based implicitly motion estimation \cite{li2016maximum}. In recent years, CNN based optical flow estimation \cite{NimaKhademiKalantari2019DeepHV} and deformable convolution \cite{chen2021hdr} are also introduced for alignment. However, existing HDR video reconstruction methods are all trained on the synthetic video dataset \cite{chen2021hdr,YitongJiang2021HDRVR,NimaKhademiKalantari2019DeepHV} and the lacked details in wrong-exposed regions heavily affects the alignment and fusion accuracy.
As stated in Sec. 1, HDR reconstruction in raw domain can benefit the reconstruction quality due to the wider bit depth. Therefore, we collect a real-world HDR video dataset with both sRGB and raw formats. In this way, we not only improve the benchmark models by retraining on our dataset, but also establish a benchmark raw video HDR reconstruction method.

\begin{figure*}
	\centering
	\includegraphics[width=0.9\linewidth]{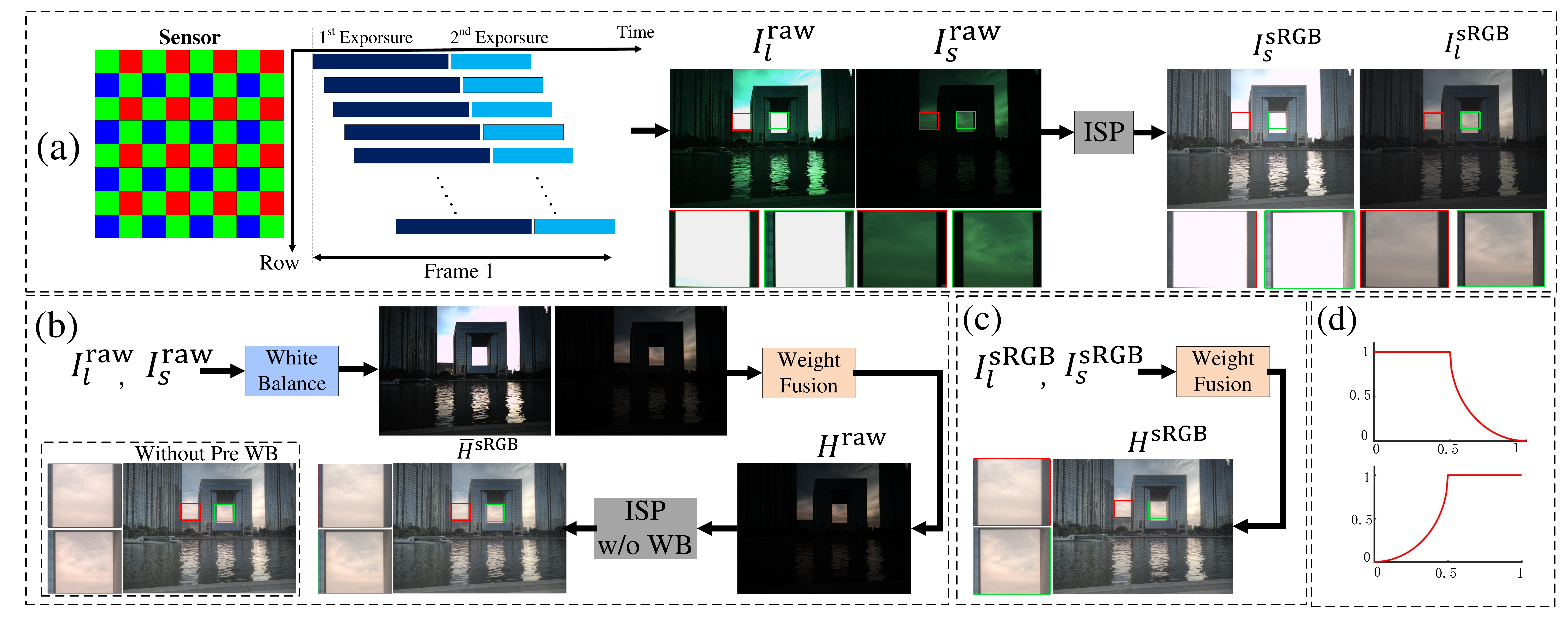}
	\vspace{-0.2cm}
	\caption{The processing pipeline of constructing LDR-HDR pairs in both raw and sRGB domains. The four subfigures represent the staggered output in both raw and sRGB domains (a), the raw HDR construction process (b), the sRGB HDR construction process (c), and the fusion curves (d). }
	\label{fig:dataset construction}
	\vspace{-0.5cm}
\end{figure*}

\section{HDR Video Dataset Construction}
The challenge in constructing HDR video dataset is capturing the corresponding HDR sequence while capturing the alternating-exposure video sequence. Different from the beam splitter strategy, we use the XiaoMi 10 Ultra phone to capture 10-bit long and short exposure images in nearly one single frame time by line interleaving technology. As Fig.~\ref{fig:dataset construction} (a) shows, the camera sensor starts the short exposure after the long exposure of each row immediately. Therefore, the long and short exposure images are overlapped in time and have very small displacement, so that they can be used to merge the HDR image. The exposure time and analog gain of each frame are automatically set according to the brightness of scenes. In this way, each frame has its own exposure time according to a certain exposure ratio. To further reduce the temporal shift between long and short exposures, we capture the video during the daytime which requires shorter exposure time.

\begin{figure}[htbp]
	\centering
	\includegraphics[width=1\linewidth]{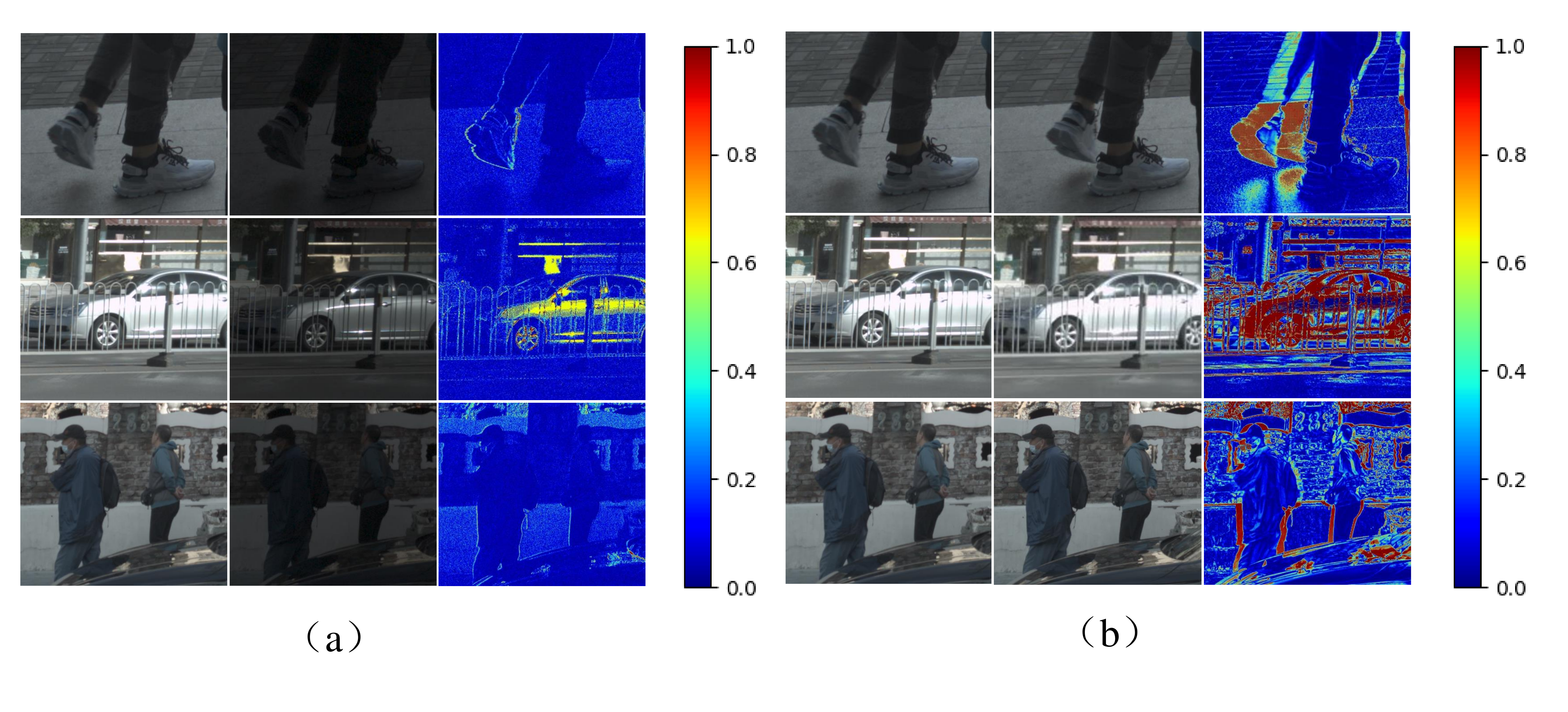}
	\caption{The comparison of displacements of staggered exposures (a) and frame-based exposures (b). For each group, the first two columns present the neighboring LDR frames (images) and the third column is the difference between the two LDR frames (images).}
	\label{fig:dispalcement}
\end{figure}

Since the short and long exposure images for each frame are nearly aligned (as shown in Fig.~\ref{fig:dispalcement} (a)), we reconstruct the corresponding HDR frame by a specific weighting function, similar to the triangular weighting~\cite{NimaKhademiKalantari2017DeepHD}, as Fig.~\ref{fig:dataset construction} (d) shows (the top (bottom) curve is the weight for long (short) exposure image). We generate the HDR ground truth in both raw ($H^{\text{raw}}$) and sRGB ($H^{\text{sRGB}}$) domains to facilitate the HDR reconstruction network training in both domains. For raw domain fusion, we first perform white balance on the raw inputs ( $I_l^{\text{raw}}$ and $I_l^{\text{raw}}$) and then perform weight fusion on the white balanced results. The fused raw HDR frame $H^{\text{raw}}$ can further go through an ISP module without white balance to transform the raw HDR to sRGB HDR (denoted as $\bar{H}^{\text{sRGB}}$). Note that, if we do not perform white balance before the fusion process, the fused result will be red in over-exposed regions. The reason is that during fusion, we scale the over-exposed regions, whose red channel values are similar to the green channel values. After fusion and white balance, the red values will be larger in the over-exposed regions. In contrast, by performing white balance first can help avoid the color cast artifacts in sRGB domain. The sRGB domain fusion is similar to that of raw domain and there is no special processing needed. In this way, we construct the LDR-HDR pairs in raw domain, denoted by $\{I_{l(s)}^{\text{raw}}, H^{\text{raw}}\}$, and the pairs in sRGB domain, denoted by $\{I_{l(s)}^{\text{sRGB}}, H^{\text{sRGB}}\}$.

\begin{figure}[htbp]
	\centering
	\includegraphics[width=1\linewidth]{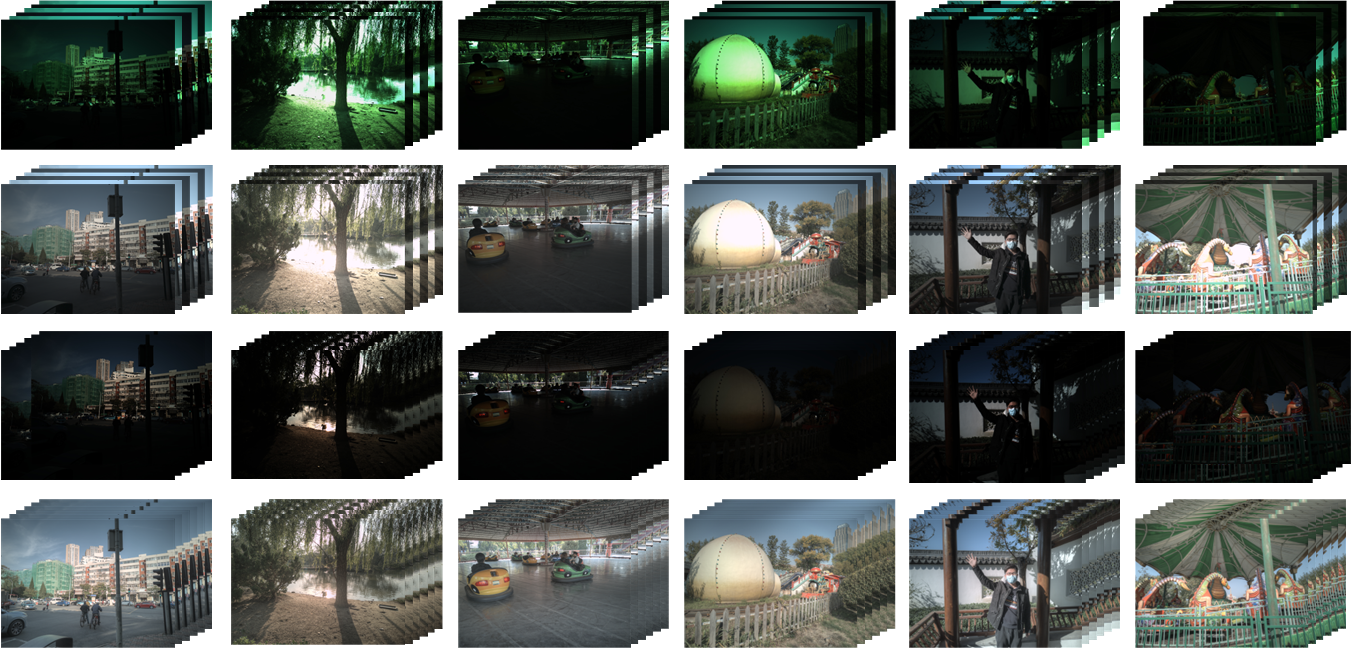}
	\caption{Exampled scenes in our dataset. From top to down, each row lists the raw LDR videos, sRGB LDR videos, raw HDR videos, and sRGB HDR videos, respectively. The sRGB HDR videos are tonemapped by photomatix for visualization.}
	\label{fig:dataset_sample}
\end{figure}

We manually remove some out-of-focus, wrong-exposed, or very fast motion videos, which will result in ghosting artifacts after weighted fusion.
Fig.~\ref{fig:dispalcement} (a) presents the displacements between the short and long exposure images in one frame time. Specifically, the displacements between the short and long exposure images are calculated by scaling them to the same exposure level in raw domain.
It can be observed that the displacements only exist in the
over-exposed regions (displayed in yellow), since the over-exposed regions cannot be scaled to the same value as those in the short exposure image. In contrast, Fig.~\ref{fig:dispalcement} (b) presents the displacements between the neighboring same exposure frames. The displacements are much larger due to the object movements and camera motions. Therefore, the constructed dataset is suitable for training and evaluation of HDR models.
We collect 85 LDR-HDR videos with alternating exposures and the scenes are various, such as urban streets, zoos, parks, and buildings. Then, we keep one exposure image for each frame alternatively to construct the input LDR sequence. All the captured videos have both raw and sRGB versions with per-frame ground truth HDR, as shown in Fig.~\ref{fig:dataset_sample}. Note that, the bit depth of each frame in our dataset is 10, and the exposure ratio between long and short exposure images varies between 1:8 to 1:9, according to the scene illumination. Therefore, the HDR video improves the dynamic range of LDR video by 18 dB. 


\setlength{\tabcolsep}{2.5pt}
\begin{table}[htbp]
	\begin{center}
		\caption{Comparison between our and other real-world datasets for video HDR with two exposures.}
		\label{table:datasets}
		\begin{tabular}{cccccc}
			\hline\noalign{\smallskip}
			Data & Resolution & Video Numbers & Frames & GT & Raw Format\\
			\noalign{\smallskip}
			\hline
			\noalign{\smallskip}
			\cite{NimaKhademiKalantari2013PatchbasedHD} & $1280 \times 720$ & 5 & 50-200 & w/o & w/o\\
			\cite{chen2021hdr} & $4096 \times 2168$ & 76 & 5-7 & central frame & w/\\
			Ours & $4000 \times 3000$ & 85 & 60 & per frame & w/\\
			\hline
		\end{tabular}
	\end{center}
\end{table}
\setlength{\tabcolsep}{1.4pt}

Table~\ref{table:datasets} summarizes the comparisons of our dataset with other real-world datasets
\cite{chen2021hdr,NimaKhademiKalantari2013PatchbasedHD}. It can be observed that our dataset is the largest and the only one that has per-frame ground truth. Compared with the dynamic dataset in ~\cite{chen2021hdr}, our dataset is more diverse in terms of scenes and actions. 

\section{Approach}

\begin{figure*}
	\centering
	\includegraphics[width=1\textwidth]{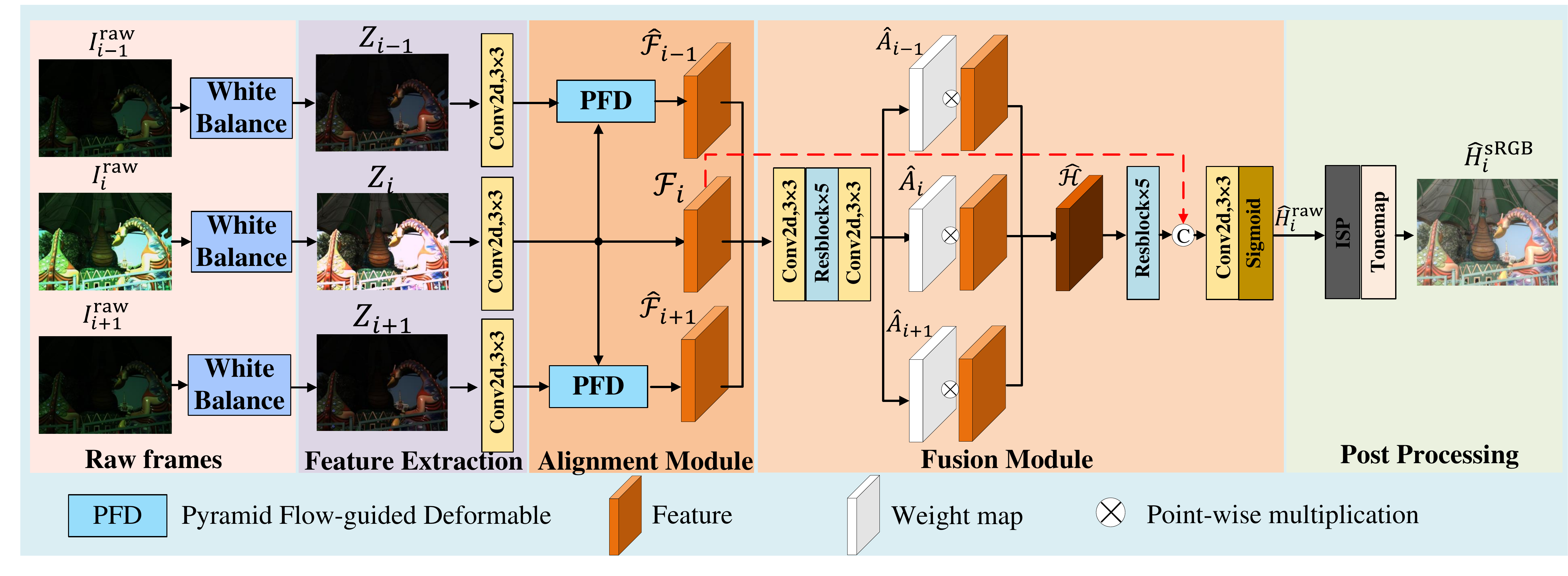}
	\caption{The architecture of the proposed Raw-HDRNet, which includes a flow-guided pyramid deformable (FPCD) for alignment and a weight predictor for fusion. The fused raw HDR result is converted to sRGB domain by an ISP module without white balance.}
	\label{fig:network structure}
	\vspace{-0.5cm}
\end{figure*}

Given a raw LDR video ${\{I_i^{\text{raw}}\mid i=1,...,n\}}$ captured with alternating exposures ${\{t_i\mid i=1,...,n\}}$, we aim to reconstruct the corresponding raw HDR video. Since the raw frames are characterized by Bayer patterns, we pack them into 4 channels, resulting $I_i^{\text{raw}}\in \mathbb{R}^{H\times W \times 4}$. Our Raw-HDRNet reconstructs the middle HDR frame $\hat{H}_i^{\text{raw}}\in \mathbb{R}^{H\times W \times 4}$ for every three frames. 
The HDR reconstruction process is fusing the well-exposed regions from different exposure frames and accurate alignment can reduce the ghosting artifacts generated by fusion. Therefore, the proposed Raw-HDRNet consists of feature extraction, alignment and fusion modules, as shown in Fig. ~\ref{fig:network structure}, which are described in the following.

\subsection{Feature Extraction}
With three alternating exposed raw LDR frames $I_{[i-1:i+1]}^{\text{raw}}$, we first perform white balance on them since our raw HDR ground truth has been white balanced. For brevity, we still utilize $I_{[i-1:i+1]}^{\text{raw}}$ to denote the white balanced result.
Since the raw inputs are in linear domain, we directly transform $I_i^{\text{raw}}$ to the HDR domain by
$\tilde{H}_i^{\text{raw}} = {I_i^{\text{raw}}}/{t_i}$. In this way, the HDR domain frames $\tilde{H}^{\text{raw}}_{[i-1:i+1]}$  are aligned in brightness, which is suitable for offsets estimation. Meanwhile, the LDR domain frames $I^{\text{raw}}_{[i-1:i+1]}$ can help indicate the under and over exposed regions, which is useful for the estimation of fusion weights. Therefore, we concatenate the $I^{\text{raw}}_i$ and $\tilde{H}^{\text{raw}}_i$ along the channel dimension, generating $Z_i\in \mathbb{R}^{H\times W \times 8}$ as the inputs.
Then a convolution layer is  applied on $Z_i$ to extract shallow features $\mathcal{F}_i$. The convolution weights are shared for the three frames.

\subsection{Alignment Module}
The pyramid cascading deformable (PCD) convolution \cite{wang2019edvr} is widely used in frame alignment. However, it is not stable when dealing with large offsets. To solve this problem, BasicVSR++ \cite{chan2021basicvsr++} introduces flow-guided deformable convolution in video super-resolution. However, it estimates the offsets in a single scale. In HDR reconstruction, large foreground motions between neighboring frames will lead to severe ghosting artifacts. Therefore, we propose a pyramid flow-guided deformable (PFD) convolution for alignment, which can address large movements better.

We use the SpyNet-triple~\cite{NimaKhademiKalantari2019DeepHV} as our flow network, which utilizes three consecutive LDR frames to estimate two optical flow maps. Since our LDR raw frames are alternating exposed, the reference and the neighboring frames have different brightness. Therefore, we first adjust the exposure of the reference frame to match the neighboring frame as $I'^{\text{raw}}_i=\text{clip}(I^{\text{raw}}_i*t_{i+1}/t_i)$. Considering the raw frames concentrate on low intensities, we further perform gamma correction
on the three inputs $[I^{\text{raw}}_{i-1},I'^{\text{raw}}_i,I^{\text{raw}}_{i+1}]$, generating $[{I}^{\text{g}}_{i-1},{I'}^{\text{g}}_i,{I}^{\text{g}}_{i+1}]$, where ${I}^{g}={I^{\text{raw}}}^{1/\gamma}$ to enlarge low intensity values. Then the three frames are used to estimate flow maps in a coarse-to-fine way, which can be represented as
\begin{equation}\label{Flow estimation}
[F^s_{i-1},F^s_{i+1}] = \mathcal{G}([{I}^{g}_{i-1},{I'}^g_i,{I}^g_{i+1}]),\ \ \ s = 1,...,5,
\end{equation}
where $\mathcal{G}$ represents the basic CNN for estimating optical flow and has the same structure in every scale $s$. $F^s_{i-1}$ is the flow map for the ${i-1}$ frame. 

\begin{figure}[htbp]
	\centering
	\includegraphics[width=0.48\textwidth]{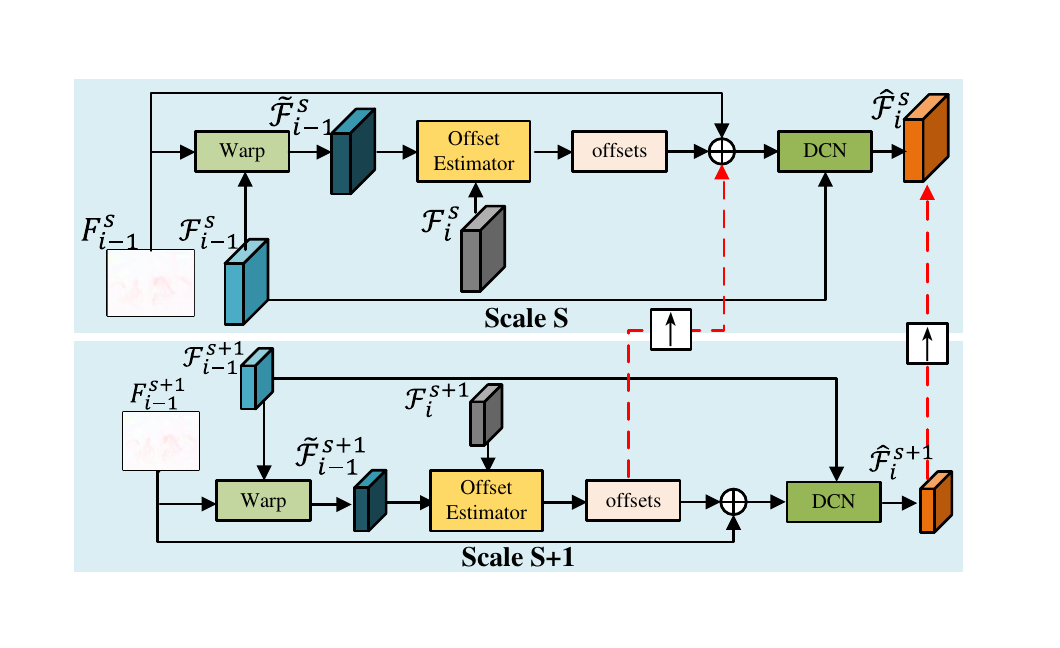}
	\vspace{-0.5cm}
	\caption{The architecture of the proposed PFD module.}
	\label{fig:FPCD structure}
	\vspace{-0.5cm}
\end{figure}

In the following, we take $\mathcal{F}_{i-1}$ as an example to demonstrate the proposed PFD process, denoted as
$\hat{\mathcal{F}}_{i-1}= f_{\text{PFD}}(\mathcal{F}_{i-1},\mathcal{F}_{i},F_{i-1})$, where
$f_{\text{PFD}}(\cdot)$ represents the PFD process.
Firstly, the PFD module downsamples the features to obtain 4-scale pyramid features,
$\mathcal{F}^{s+1}_{i}= (\mathcal{F}^{s}_{i})^{{\downarrow}2}(s = 1,2,3),$
where $(\cdot)^{{\downarrow}2}$ represents the convolution with stride of 2. We warp $\mathcal{F}^{s}_{i-1}$ with $F^s_{i-1}$, generating $\tilde{\mathcal{F}}^{s}_{i-1}$.
Then the warped features are concatenated with the reference feature to compute the offsets:
\begin{equation}\label{offset estimation}
O_{i-1}^s = \mathcal{C}([\tilde{\mathcal{F}}^{s}_{i-1},\mathcal{F}_i]),\ \ \ s = 1,...,4,
\end{equation}
where $\mathcal{C}$ denotes a stack of convolutions. The offsets $O_{i-1}^s$ plus the original optical flow $F_{i-1}^s$ resulting the final offsets for deformable convolution based alignment, which can be described as
\begin{equation}
\hat{\mathcal{F}}^s_{i-1} = \mathcal{D}(\mathcal{F}^s_{i-1},O_{i-1}^s+F_{i-1}^s),\ \ \ s = 4,
\end{equation}
\begin{equation}
\begin{split}
\hat{\mathcal{F}}^{s}_{i-1} &= \text{Conv}([\mathcal{D}(\mathcal{F}^{s}_{i-1},O_{i-1}^{s}+F_{i-1}^{s}\\
&+(O_{i-1}^{s+1})^{{\uparrow}2}),(\hat{\mathcal{F}}^{s+1}_{i-1})^{{\uparrow}2}]),\ \ \ s = 1,2,3,
\end{split}
\end{equation}
where $\mathcal{D}$ denotes the deformable convolution. For simplicity, we omit the description of modulation masks in deformable convolution. The $(\cdot)^{{\uparrow}2}$ refers to bilinear upsampling. For a finer scale, the offsets are refined by the flow map and its previous scale offsets. Then the aligned features are  concatenated with the upsampled features, which are then fused together via a convolution layer, generating the aligned result at the corresponding scale. After the pyramid alignment, a cascading DCN is performed in the finest scale to refine the aligned feature without the guidance of flow, i.e., $O_{i-1} = \mathcal{C}([\hat{\mathcal{F}}_{i-1}^1,\mathcal{F}^1_i])$ and $\hat{\mathcal{F}}_{i-1} = \mathcal{D}(\hat{\mathcal{F}}^1_{i-1},O_{i-1})$. The cascading deformable convolution helps refine the result of the previous flow-guided deformable alignment. In this way, we obtain the aligned features $\hat{\mathcal{F}}_{i-1}$ and $\hat{\mathcal{F}}_{i+1}$  for the neighboring frames.

\subsection{Fusion Module}
We utilize the fusion module to merge the aligned features together. Specifically, we utilize five ResBlocks to calculate the weighting coefficients $A_{[i-1:i+1]}= f_\mathcal{\text{Fusion}}(\hat{\mathcal{F}}_{[i-1:i+1]})$. Then, the coefficients are normalized via $\hat{A}_i = \frac{{A_{i}}}{\sum_{j=i-1}^{j=i+1} {A_{j}}}$. Finally, the fused feature map $\hat{\mathcal{H}}$ is obtained by $\hat{\mathcal{H}}=\sum_{j=i-1}^{i+1}\hat{A}_j\hat{\mathcal{F}}_{j}$. Generally, the well aligned and exposed regions are given large weights, and the wrong-exposed or misaligned regions are given small weights. Hereafter, $\hat{\mathcal{H}}$ goes through the following ResBlocks. To avoid details lost, we further introduce a skip connection to make $\mathcal{F}_i$ be concatenated with the features.
Then, after the sigmoid layer, we generate the raw HDR frame $\hat{H}^{\text{raw}}_i$. Note that, users can utilize different ISP (without white balance) to process the generated raw HDR frames. In this work, we utilize a traditional ISP to convert the raw HDR frame into sRGB frame, denoted by $\hat{H}^{\text{sRGB}}_i$.

\subsection{Loss Functions}
Since the HDR videos are usually observed in tonemapped domains, following \cite{NimaKhademiKalantari2017DeepHD}, we calculate the loss in the $\mu$-law tonemapped space. The tonemapped image $\hat{T}_i^{\text{raw}}$ is calculated by $\hat{T}^{\text{raw}}_i=\frac{\text{log}(1+\mu \hat{H}^{\text{raw}}_i)}{\text{log}(1+\mu)}$, where $\mu$ is the compression factor and it is set to 5000. The ground truth HDR image is also tonemapped in the same way. We utilize the $\ell1$ norm to calculate the loss as $\mathcal{L}=\|T^{\text{raw}}_i-\hat{T}^{\text{raw}}_i\|_1$, where ${T}^{\text{raw}}_i$ is the ground truth.

\section{Experiments}

\newsavebox\CBox
\def\textBF#1{\sbox\CBox{#1}\resizebox{\wd\CBox}{\ht\CBox}{\textbf{#1}}}
\setlength{\tabcolsep}{1pt}
\begin{table*}[htbp]
	\begin{center}
		\centering
		\caption{Comparison of HDR reconstruction results on three test sets with different training sets, i.e., the synthetic dataset and the combination of synthetic and our datasets (denoted by +). For each method, the best results are highlighted in bold.}
		\renewcommand\arraystretch{1}
		\label{tab:test results}
		\resizebox{\textwidth}{!}{
			\normalsize
			\begin{tabular}{l*{3}{c}*{3}{c}*{3}{c}}
				\toprule[1.5pt]
				& \multicolumn{3}{c}{\large Synthetic dataset}
				& \multicolumn{3}{c}{\large Real-world dataset \cite{chen2021hdr}}
				& \multicolumn{3}{c}{\large Our dataset}  \\
				Method
				& PSNR-$\mu$   &  HDR-VDP2 &  HDR-VQM
				& PSNR-$\mu$    &  HDR-VDP2 &  HDR-VQM
				& PSNR-$\mu$    &  HDR-VDP2 &  HDR-VQM\\
				
				\midrule
				Yan19~\cite{QingsenYan2019AttentionGuidedNF}
				& 38.72 & 68.10 & 76.63 & 43.15 & \textBF{77.79} & 78.92 & 37.69 & 57.64 & 93.52 \\
				Yan19+
				&\textBF{39.20} & \textBF{68.78} & \textBF{79.64} & \textBF{45.11} & 76.84 & \textBF{88.90} & \textBF{44.02} & \textBF{61.09} & \textBF{93.74} \\
				\midrule
				
				Kalantari19 ~\cite{NimaKhademiKalantari2019DeepHV}
				& 37.48 & 70.67 & \textBF{84.57} & 44.72 & 77.91 & 87.16 & 43.94 & 61.83 & 95.10 \\
				Kalantari19+
				& \textBF{39.12}  & \textBF{71.07} & 84.53 & \textBF{44.85} & \textBF{78.19} & \textBF{87.69} &  \textBF{45.32} &  \textBF{62.47} & \textBF{95.50}\\
				\midrule

				Chen21~\cite{chen2021hdr}& 40.34  & 71.79 & \textBF{85.71} & 45.46 & 79.09 & \textBF{87.40} & 44.26 & 62.67 & 96.11 \\
				Chen21+& \textBF{41.06} & \textBF{71.89} & 85.04 & \textBF{45.60} & \textBF{79.28} & 87.11 & \textBF{44.47} & \textBF{63.80} &\textBF{96.60}\\
				\bottomrule[1.5pt]{\smallskip}
			\end{tabular}
		}
		\vspace{-2em}
	\end{center}
\end{table*}


In this section, we first evaluate the benefits of introducing our dataset by training with sRGB LDR inputs and then verify the effectiveness of proposed Raw-HDRNet by training with raw inputs.
We use PSNR-$\mu$ to evaluate the results in the $\mu$-law tonemapped domain. In addition, we also adopt HDR-VDP-2~\cite{RafalMantiuk2011HDRVDP2AC} and HDR-VQM~\cite{narwaria2015hdr} to evaluate the visual quality.

\subsection{sRGB Domain HDR Reconstruction}

We evaluate the benefits of introducing our dataset by retraining three benchmark networks \cite{chen2021hdr,NimaKhademiKalantari2019DeepHV,QingsenYan2019AttentionGuidedNF} on the mixed dataset, i.e., the synthetic dataset \cite{JanFroehlich2014CreatingCW,kronander2014unified,TianfanXue2019VideoEW} and our dataset. The baseline results are generated by retraining the three networks on the synthetic dataset only.
The synthetic dataset includes 21 HDR videos from ~\cite{JanFroehlich2014CreatingCW,kronander2014unified} and video clips from Vimeo90K \cite{TianfanXue2019VideoEW}, where the LDR sequences are synthesized from ground truth sequences with different scalings and estimated CRF curves. 
Our sRGB dataset contains 85 videos, and we randomly select 79 of them for training and the rest for testing. We follow \cite{NimaKhademiKalantari2019DeepHV} for data augmentation. Specifically, we add Gaussian noise, with variance varying from 1e-3 to 3e-3, on the linearized LDR frames to simulate the noise in real captured scenes. We also perturb the tone of the reference frame with a gamma function ($ \text{exp}(d),d\in[-0.7,0.7]$) to simulate the camera response function for the synthetic dataset. We crop patches of size $256\times256$ for training. Since the resolution of image in our dataset is large ($4000\times3000$), we only select the patches which have large movements for training. We retrain Yan19~\cite{QingsenYan2019AttentionGuidedNF} and Chen21~\cite{chen2021hdr} with their released codes. As there is no public code for Kalantari19~\cite{NimaKhademiKalantari2019DeepHV}, we re-implement their network according to their settings claimed in the paper.

\begin{figure*}[htbp]
	\centering
	\includegraphics[width=1\textwidth]{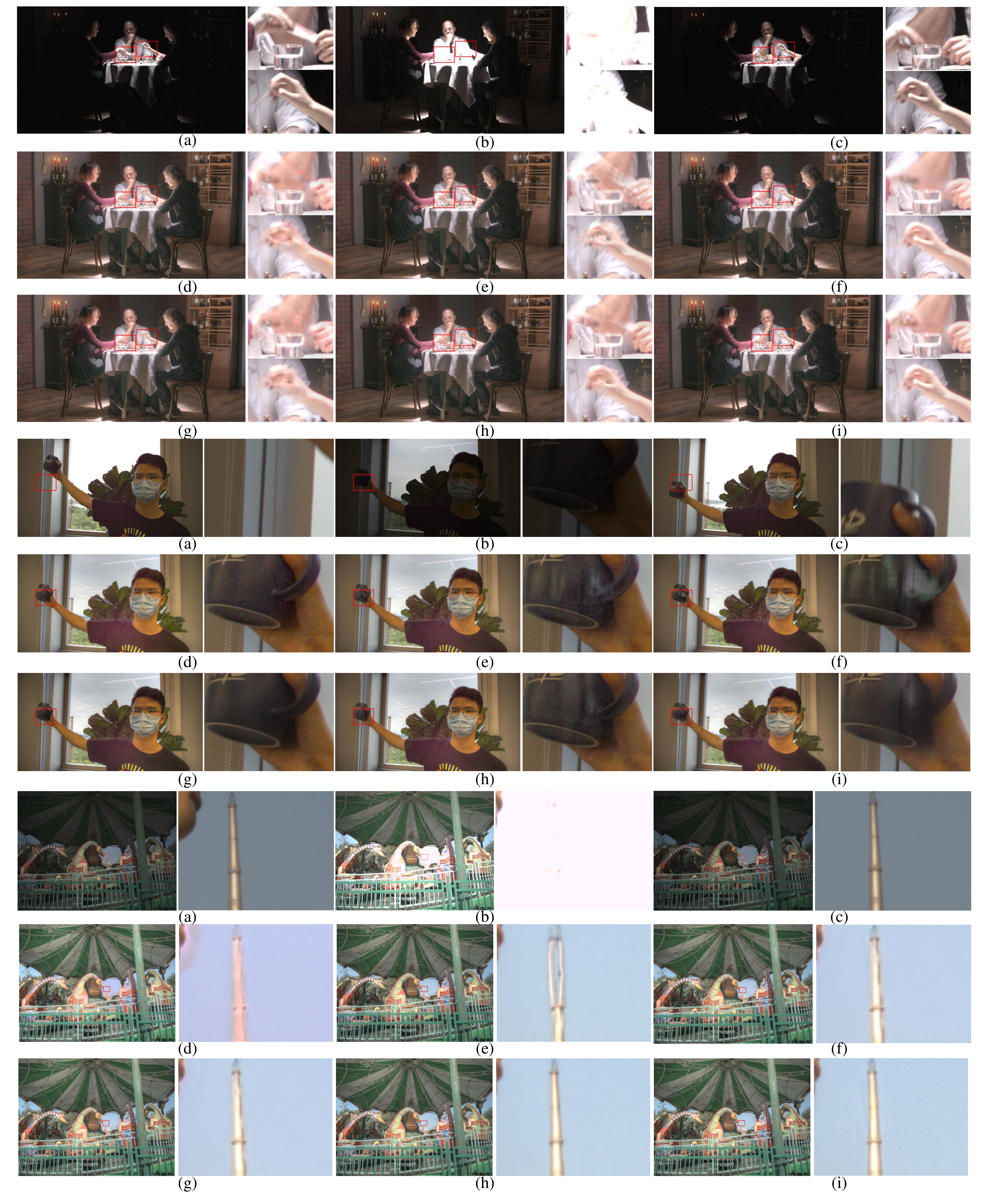}
	\caption{Visual comparisons for different methods trained with different datasets. For each group, the first row (a-c) represents the LDR inputs. The second and third row are the results generated by models trained with only synthetic dataset and  mixed dataset, respectively. From left to right, the results are generated by Yan19 (d,g), Kalantari19 (e,h), and Chen21 (f,i). For each image, we zoom in highlighted areas for better observation. The three groups are from three different test sets.}
	\label{fig:rgb_com}
\end{figure*}

We evaluate the performance of the three networks on two benchmark test sets, i.e., the synthesized test sets (POKER FULLSHOT and CAROUSEL FIREWORKS) from \cite{JanFroehlich2014CreatingCW} and the real-world test set (75 scenes, where each scene contains one HDR frame) from  \cite{chen2021hdr}, and on our real-world test set (six videos, where we sample 10 frames for each video). 
The comparison results are shown in Table~\ref{tab:test results}. It can be observed that introducing our dataset for training, all the three methods have achieved better performances. Specifically, by introducing our real-world dataset for training, the performance on synthetic dataset is also improved. Directly utilizing the model trained with synthesized dataset cannot work well on our dataset, which demonstrates that the generalization ability of the synthesized dataset is limited. 
By retraining with mixed dataset, the generalization on the third dataset (i.e., the real-world dataset from \cite{chen2021hdr}) is also improved. The above experimental results show that the proposed dataset can effectively improve the performance of existing models and has good generalization.

\setlength{\tabcolsep}{1 pt}
\begin{table*}[!htbp]
	\begin{center}
		\centering
		\caption{HDR reconstruction results on our test set using models trained with our dataset in sRGB domain and raw domain, respectively. The results are evaluated in sRGB domain.}
		\renewcommand\arraystretch{1}
		\label{tab:raw results}
		\normalsize
		\begin{tabular}{l*{3}{c}*{3}{c}}
			\toprule[1.5pt]
			& \multicolumn{3}{c}{\large sRGB domain}
			& \multicolumn{3}{c}{\large Raw domain}\\
			Method
			& PSNR-$\mu$   &  HDR-VDP2 &  HDR-VQM
			& PSNR-$\mu$    &  HDR-VDP2 &  HDR-VQM\\
			
			\midrule
			Yan19~\cite{QingsenYan2019AttentionGuidedNF} &  44.58  & 61.28  & 94.69  & 45.64 & 63.76  &95.59 \\
			\midrule
			Kalantari19~\cite{NimaKhademiKalantari2019DeepHV}& 45.03 & 61.82 & 95.40 & 46.13 & \textBF{63.93} & 95.60 \\
			\midrule
			Chen21~\cite{chen2021hdr}& 44.38  & 63.05 & 96.01 & 45.64 & 62.28 & 95.32\\
			\midrule
			Ours  &  45.02 & 61.44 &  95.61 & \textBF{46.75} & \textBF{63.93} & \textBF{96.12}\\
			\bottomrule[1.5pt]{\smallskip}
		\end{tabular}
	\end{center}
\end{table*}

 Fig.~\ref{fig:rgb_com} presents the visual comparison results for three different methods with different training sets, and more results are provided in the supplementary file. 
 %
The first group shows the test results of different HDR reconstruction methods on one scene of the synthetic dataset (video POKER FULLSHOT) and the reference is a long exposure frame. It can be observed the results in the second row show obvious ghosting artifacts and severe blur when there is movement in the highlighted areas. In contrast, the results of all methods in the third row have less ghosting artifacts and the reconstructed details are much clean. The second group shows the results on the real multi-frame dataset~\cite{chen2021hdr}and the reference is a short exposure frame. In high-contrast indoor scenes, all methods generate ghost artifacts because of the alignment and fusion errors. It can be observed that by introducing our dataset for training, the ghosting artifacts can be reduced and the noise can be well removed. The main reason is that our dataset contains various scenes and motions, which is a good complementary of the synthetic dataset. 
%
%
The third group shows the test results on our proposed dataset. The model trained with only synthetic dataset leads to color bias problems and detail loss. The synthetic dataset can not estimate the CRF curve accurately, which affects the linearization during HDR reconstruction. In contrast, our proposed dataset has accurate linear properties and benefits supervised training for HDR reconstruction.

\subsection{Raw Domain HDR Reconstruction}

\subsubsection {Implementation Details}
Since raw frame is linear, there is no need to perform tone perturbations. 
Our network is optimized with Adam. The initial learning rate is set to of 1e-4, which is decreased to 1e-5 after 30 epochs. The whole network converges after 50 epochs. The batch size is 16 and the patch size is $256\times256$. The proposed method is implemented in Pytorch and trained with an NVIDIA 3090 GPU.

\subsubsection {Comparison with State-of-the-arts}

\begin{figure*}[!htbp]
	\centering 
	\includegraphics[width=1\linewidth]{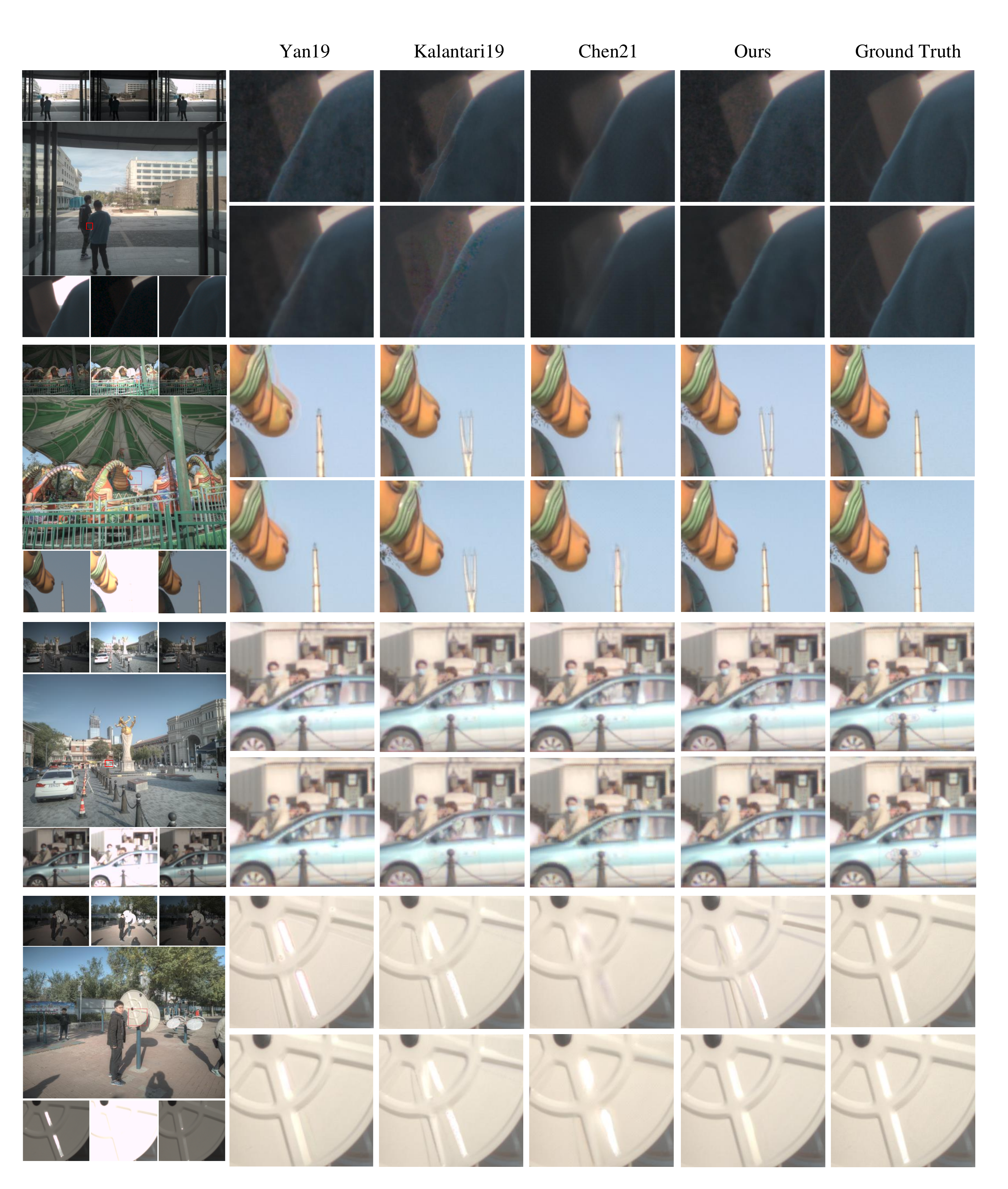}
	\caption{Comparison results of different methods trained with sRGB domain LDR-HDR pairs (top row in each scene) and raw domain LDR-HDR pairs (bottom row in each scene). The left column is the LDR inputs (patches) and our full-resolution HDR result with raw inputs.}
	\label{raw_result}
\end{figure*}


We evaluate the superiority of utilizing raw inputs over the sRGB inputs for HDR reconstruction by training our network and three benchmark networks on our dataset. For all the methods, we report two results, where one is generated with sRGB inputs and the other is generated with raw inputs. We revise the codes of \cite{chen2021hdr,NimaKhademiKalantari2019DeepHV,QingsenYan2019AttentionGuidedNF} to make them adapt to raw inputs. Specifically, we make their first convolution layer adapt to four channel inputs. Our method and \cite{chen2021hdr} includes flow networks, and we initialize them with the pretrained flow parameters on synthesized HDR datasets. Similar to our method, the loss functions of \cite{NimaKhademiKalantari2019DeepHV,QingsenYan2019AttentionGuidedNF} are the $\ell1$ loss in the $\mu$-law tonemapped domain. For \cite{chen2021hdr}, since it includes VGG loss in the sRGB domain, we further introduce an ISP module to the raw HDR output and calculate the VGG loss in the sRGB domain.

As shown in Table \ref{tab:raw results} \footnote{Note that, the results in this Table are different from those in Table \ref{tab:test results}. The models in Table \ref{tab:test results} are trained with synthetic and mixed dataset (our dataset + synthetic dataset). In contrast, the models in \ref{tab:raw results} are trained with only our dataset. We did not utilize the mixed dataset in this experiment since the synthetic dataset does not have raw inputs.}, the results of most methods in raw domain are better than their corresponding results in sRGB domain. Compared with the results in sRGB domain, Yan's method improves PSNR-$\mu$ by 1.06 dB, HDR-VDP by 2.48 and HDR-VQM by 0.9 in raw domain. As for Kalantari's method, the PSNR-$\mu$ is improved by 1.10 dB after training in the raw domain. This demonstrates that performing HDR fusion in raw domain is beneficial due to the wider bit depth and linearity of raw inputs. However, although Chen's method increases the PSNR-$\mu$ by 1.26 dB, HDR-VDP2 and HDR-VQM are decreased because their heavy denosing ability in raw domain makes image details smooth and reduces the scene brightness. 
Our method achieves the best results in terms of PSNR-$\mu$ and HDR-VQM, which verifies that our method can realize high-quality HDR reconstruction and improve video consistency (HDR-VQM can measure the consistency of video frames). Therefore, our method is a strong baseline for raw HDR reconstruction.

Fig.~\ref{raw_result} presents the visual comparison results of different methods trained in sRGB domain and raw domain respectively, and more results are provided in the supplementary file. The first row of each scene is the results generated by methods trained in sRGB domain, and the second row of each scene is the results generated by methods trained in raw domain. The reconstruction results in raw domain have been converted to sRGB domain after ISP processing, and all HDR images are tonemapped by Photomatix pro for observation.

As can be seen from the figure, performing HDR reconstruction in raw domain can reconstruct higher quality results, including fewer artifacts, less noise, and more details. Yan's method is prone to produce ghosts in sRGB domain but greatly reduced in raw domain. In addition, their denoising ability is also improved. Kalantari's method introduces alignment artifacts in both sRGB domain and raw domain, due to complex lighting conditions and object motion occlusion, which indicates that optical flow can not achieve high-quality image alignment. Chen's method can reduce ghost artifacts and recover the over-exposed details in raw domain. However, this method tends to over-smooth details and can not completely reconstruct the clear appearance of objects. Finally, our proposed method also leads to ghost artifacts in sRGB domain due to the limited bit width, but achieves the best performance in raw domain through accurate feature alignment and fusion. The results of our method in raw domain have fewer artifacts and restore complete details in highlight areas.

\subsubsection {Ablation Study}

We perform ablation experiments to verify the effectiveness of HDR reconstruction in raw domain and our proposed alignment module. As shown in Table \ref{tab:ablation}, "w/o" means not included and "Raw" means raw domain input. "PFD" means the proposed pyramid flow-guided deformable convolution alignment module, and "w/o Raw" means the network is trained with sRGB domain data. "W/O PFD" means to replace the PFD module with the PCD module~\cite{wang2019edvr}.

\begin{table}[htbp]
	\setlength{\tabcolsep}{4pt}
	\begin{center}
		\centering
		\caption{Ablation study on our dataset.The best results are highlighted in bold.}
		\label{tab:ablation}
		\small
		\renewcommand\arraystretch{1.25}
		\begin{tabular}{cccccc}
			\toprule[1.5pt]
			Method & Raw  &  PFD  & PSNR-$\mu$   &  HDR-VDP2 &  HDR-VQM\\
			\midrule
			w/o Raw &  \XSolid   & \Checkmark  & 45.02 & 61.44 & 95.61\\
			w/o PFD &  \Checkmark   & \XSolid   & 46.18 & 63.75 & 95.88\\
			Full Structure  &  \Checkmark & \Checkmark & \textBF{46.75} & \textBF{63.93} & \textBF{96.12}\\
			\bottomrule[1.5pt]{\smallskip}
		\end{tabular}
	\end{center}
\end{table}

As can be seen from the table, if we process in sRGB domain, the PSNR-$\mu$ will drop by 1.73dB, HDR-VDP2 decreased by 2.49, and HDR-VQM decreased by 0.51, indicating the necessity of training in raw domain. If we replace the PFD with PCD module, the results degrade by 0.57dB, which indicates the proposed PFD module can learn accurate alignment and improve the quality of the final HDR results.

\subsubsection {Complexity Comparison}

Table \ref{tab:Complexity} lists the computation costs and inference times of our method and compared methods by evaluating with an NVIDIA RTX 3090 GPU.
It can be observed that our method is heavier than \cite{QingsenYan2019AttentionGuidedNF} \cite{NimaKhademiKalantari2019DeepHV}, but is lighter than \cite{chen2021hdr} in terms of TFLOPs. In the future, we will optimize our network structure to reduce computing complexity.

\setlength{\tabcolsep}{2pt}
\begin{table}[htbp]
	\begin{center}
		\centering
		\caption{Comparisons of TFLOPs, inference time, and parameters for a $1920 \times 1080$ input.}
		{\smallskip}
		\renewcommand\arraystretch{1.25}
		\label{tab:Complexity}
		\begin{tabular}{ccccc}
			\toprule[1.5pt]
			Methods & Yan19 \cite{QingsenYan2019AttentionGuidedNF}  &  Kalantari19 \cite{NimaKhademiKalantari2019DeepHV} &  Chen21 \cite{chen2021hdr}& Ours\\
			\midrule
			TFLOPs (T) &  2.61  & 3.17  & 7.27  & 5.97 \\
			Inference Time (s) & 0.03 & 0.07  & 0.55  & 0.77  \\
			Parameters (M) & 1.32  & 10.38 & 6.44 & 4.43\\
			\bottomrule[1.5pt]{\smallskip}
		\end{tabular}
	\end{center}
\end{table}

\section{Conclusions}
In this work, we construct an LDR-HDR video dataset in both raw and sRGB domains. We propose a suitable method to construct the raw HDR ground truth by performing white balance before the fusion process. Correspondingly, we propose a Raw-HDRNet for HDR reconstruction in the raw domain. We design a pyramid-flow guided deformable convolution for alignment, where the flow is calculated in gamma corrected LDR raw frames, and the deformable offset is calculated with the features extracted from both LDR and HDR raw inputs. Experimental results demonstrate the effectiness of the proposed dataset and Raw-HDRNet. 

Note that since our dataset is captured in daytime, the model trained with only our dataset may cannot deal with the noise in dark regions well. Since it is difficult to capture well-aligned short and long exposure frames to construct ground truth for night scenes, in the future, we would like to unprocess the synthesized night scenes to raw domain to augment our dataset. In addition, the LDR videos of our constructed dataset only contain 10 bits, which limits the final dynamic range of our HDR video. This can be improved by capturing with more professional devices. Our dataset construction method and Raw-HDRNet are also suitable for videos captured with new devices. 

\section*{Acknowledgment}

The authors would like to thank Jixin Zhao for his help during dataset construction.

\ifCLASSOPTIONcaptionsoff
  \newpage
\fi

\bibliographystyle{IEEEtran}
\bibliography{reference}{}
%


%








\end{document}